\documentclass[manuscript,screen,nonacm]{acmart}

\setcopyright{none}
\makeatletter
\renewcommand{\@formatdoi}[1]{\ignorespaces}
\makeatother

\title{GazeCopilot: Evaluating Novel Gaze-Informed Prompting for AI-Supported Code Comprehension and Readability}

\author{Yasmine Elfares}
\affiliation{%
  \institution{University of Glasgow}
  \country{United Kingdom}
}
\email{y.elfares.1@research.gla.ac.uk}

\author{Gül \c {C}alikli}
\affiliation{%
  \institution{University of Glasgow}
  \country{United Kingdom}
}
\email{HandanGul.Calikli@glasgow.ac.uk}

\author{Mohamed Khamis}
\affiliation{%
  \institution{University of Glasgow}
  \country{United Kingdom}
}
\email{Mohamed.Khamis@glasgow.ac.uk}

\renewcommand{\shortauthors}{Elfares et al.}

\usepackage{graphicx}
\usepackage{xspace}
\usepackage{tabularx} 
\usepackage[most]{tcolorbox}
\usepackage{enumitem}
\usepackage{adjustbox} 
\usepackage{tabularray}
\usepackage[table]{xcolor}
\usepackage{tikz}
\usetikzlibrary{positioning, shapes.geometric, fit, arrows.meta}
\usepackage{array}
\usepackage{colortbl}
\usepackage{float}
\usepackage{subcaption}
\usepackage{multirow}

\newcommand{\standard}{\textsc{Standard Copilot}\xspace}
\newcommand{\gaze}{\textsc{Real-time GazeCopilot}\xspace}
\newcommand{\gazeAndText}{\textsc{Pre-set GazeCopilot}\xspace}
\newcommand{\interactionMode}{\textsc{Interaction Mode}\xspace}
\newcommand{\interactionModes}{\textsc{Interaction Modes}\xspace}
\newcommand{\systemname}{\textsc{GazeCopilot}\xspace}

\newboolean{showcomments}
\setboolean{showcomments}{true}
\usepackage{ifthen}

\ifthenelse{\boolean{showcomments}}
{
	\newcommand{\nb}[3]{%
		{\colorbox{#2}{\bfseries\sffamily\scriptsize\textcolor{white}{#1}}}%
		{\textcolor{#2}{$\blacktriangleright$\textsf\small{#3}$\blacktriangleleft$}}}
}{
	\newcommand{\nb}[3]{}
}

\newcommand{\GC}[1]{\nb{Comment:}{orange}{#1}}

\definecolor{gray05}{gray}{.95}
\definecolor{gray20}{gray}{.8}

\newlength\Linewidth
\def\findlength{\setlength\Linewidth\linewidth
	\addtolength\Linewidth{-4\fboxrule}
	\addtolength\Linewidth{-3\fboxsep}
}

\newcommand{\rb}[1]{
	\vspace{0.1cm}
	\begin{tcolorbox}[colback=gray!05,
		colframe=black,
		width=\columnwidth,
		arc=3mm, auto outer arc,
		boxrule=0.5pt,
		]
		#1
	\end{tcolorbox}
}

\newcounter{Finding}
\stepcounter{Finding}
\newcommand{\roundedbox}[1]{
	\rb{
		\noindent
		\textit{\textbf{Finding \theFinding}. #1}
	}
	\stepcounter{Finding}
}

\begin{document}

\begin{abstract}
AI-powered coding assistants, like GitHub Copilot, are increasingly used to boost developers' productivity. 
However, their output quality hinges on the contextual richness of the prompts.
Meanwhile, gaze behaviour carries rich cognitive information, providing insights into how developers process code.
We leverage this in \gaze, a novel approach that refines prompts using real-time gaze data to improve code comprehension and readability by integrating gaze metrics, like fixation patterns and pupil dilation, into prompts to adapt suggestions to developers' cognitive states.
In a controlled lab study with 25 developers, we evaluated \gaze against two baselines: \standard, which relies on text prompts provided by developers, and \gazeAndText, which uses a hard-coded prompt that assumes developers' gaze metrics indicate they are struggling with all aspects of the code, allowing us to assess the impact of leveraging the developer's personal real-time gaze data.
Our results show that prompts dynamically generated using developers' real-time gaze data significantly improve code comprehension accuracy, reduce comprehension time, and improve perceived readability compared to \standard. Our \gaze approach selectively refactors only code aspects where gaze data indicate difficulty, outperforming the overgeneralized refactoring done by \gazeAndText by avoiding revising code the developer already understands.
\textbf{Data and Materials:} \href{https://figshare.com/s/f08fec00e7b6e6badae1}{https://figshare.com/s/f08fec00e7b6e6badae1}
\end{abstract}

\maketitle


\section{Introduction}

The rapid advancements in artificial intelligence (AI) have significantly transformed software development, with AI-assisted coding tools revolutionizing the way programmers write and interact with code. These AI-powered assistants generate code suggestions based on natural language prompts, improving developer productivity and efficiency. A key factor influencing the effectiveness of these tools is prompt engineering, as the quality of the prompts directly affects the accuracy and usefulness of the LLM generated code.
It is equally important to not only generate functional but also readable and comprehensible code. Poor code readability can lead to increased cognitive load, difficulty in maintenance, and a higher likelihood of introducing errors~\cite{politowski2020large}.

Despite the efficiency gains provided by AI-assisted coding tools, two major problems persist. First, developers, especially novices, often struggle to craft effective prompts that produce clear and readable code \cite{nguyen2024beginning, liang2024large, feldman2024non}. Ineffective prompts can lead to ambiguous or overly complex AI-generated code \cite{fagadau2024analyzing}, making it even harder for beginners to comprehend and refine their code, as they may not know how to improve it or identify specific issues to fix. Second, current AI models primarily optimize for functional correctness but do not explicitly consider how programmers cognitively process the generated code \cite{kou2024large}. This gap results in AI-generated code that, while mostly syntactically correct, may not be easily readable, maintainable, or aligned with programmers' expectations \cite{al2022readable}. 

Addressing these challenges is important as readable code is crucial for software maintainability, collaboration, and debugging. Code that is difficult to comprehend can slow down development, introduce errors, and increase technical debt \cite{tornhill2022code}. As AI-assisted programming becomes more prevalent, ensuring that generated code meets readability standards is essential for fostering effective human-AI collaboration.
Secondly, incorporating human cognitive insights and physiological data into AI-driven coding can potentially improve collaborative human-AI programming. 


To this end, we propose \gaze, a novel approach for integrating real-time eye-tracking data into prompt design. As gaze data reflects developers' comprehension and cognitive load through fixations, pupil dilation, and other gaze metrics \cite{aljehane2023studying}, integrating it into prompts provides LLMs with richer context about the developer’s mental state and actual understanding of the code. 
This has the potential to bridge the gap between developers' cognitive processing and prompt engineering, making AI-assisted coding more accessible and effective, especially for those who struggle to write effective prompts.

We then report on an empirical evaluation of \gaze in a controlled lab study where 25 participants read three messy code snippets while their gaze data was recorded, then improved each snippet using one of three AI-assistants: \textbf{(i)} \standard (baseline), where participants crafted their own textual prompts, \textbf{(ii)} \gaze, where the prompt was automatically generated based on the developer's \textbf{\textit{real-time}} gaze data that indicated comprehension difficulty, and \textbf{(iii)} \gazeAndText, where we used a \textbf{\textit{fixed}} prompt to refactor the code. The \gazeAndText treatment serves as a second control condition (besides \standard) and assumes that all of the developer's gaze metrics provided indications of high cognitive load and that the code required extensive refactoring without considering developers' actual real-time gaze data. 
This condition allowed us to isolate the effect of adding generic gaze-related information from that of developers’ actual real-time gaze behaviour.
To prevent systematic biases, we counterbalanced the order of AI assistant and code snippet assignments across participants.
We measured participants’ comprehension of the refactored code, their perceived code readability, and their experience in terms of perceived control, agency, and code ownership.





While our work is not the first to study gaze behaviour during programming, \gaze is, to the best of our knowledge, the first system that integrates physiological data to improve prompts in software development contexts. 
Unlike existing AI-assisted coding solutions that rely solely on textual input, our approach incorporates real-time cognitive feedback from programmers as interpreted from their gaze data. 
This enables a novel interaction paradigm where AI outputs are shaped by insights about the developer's cognitive state.

Our results show that integrating real-time gaze data into prompt design significantly improves code comprehension and perceived readability. 
Participants rated \gaze-refactored code as statistically significantly more readable than \standard-refactored code ($p_{\text{adj}} = .001$, mean rank = 1.44 vs. 2.60). \gaze also yielded significantly higher comprehension accuracy than \standard ($p = .0122$) and than the \gazeAndText ($p = .0137$).

Our proposed approach selectively refactors code aspects associated with comprehension difficulty inferred from developers’ gaze behaviour. By focusing only on these code aspects, our method reduces the overhead of revising code that developers already understand, avoiding excessive refactoring with a pre-set prompt.
This aligns with prior work showing that refactoring can be costly, risky, and context-dependent, and its benefits vary across different code regions \cite{kim2014empirical}, and that ad hoc or indiscriminate refactoring can be counterproductive \cite{Buschmann2011}. 
These findings highlight the potential of real-time gaze data to meaningfully enhance prompt quality and improve human-AI collaboration in software development.


This research contributes to the broader goal of making AI tools more user-centric, improving both productivity and code quality in software development environments.
Bridging cognitive feedback and AI prompt engineering opens new possibilities for more human-aware, adaptive coding assistants that go beyond purely text-based interactions. This could make AI-assisted programming more accessible, reduce cognitive load, and improve code quality and maintainability at scale. 
Beyond programming, it also demonstrates a method for integrating real-time physiological signals into AI systems, paving the way for more intuitive, context-aware human-AI collaborations.


\section{Related Work}
Our work is situated at the intersection of eye-tracking and AI-assisted programming. While prior research has explored each of these areas independently, integrating real-time gaze signals into LLM prompts remains underexplored.

\smallskip
\noindent \textbf{AI-Assisted Coding and Prompt Engineering.} AI coding tools, such as GitHub Copilot, can have a significant impact on aspects of developer productivity, including a decrease in development task time~\cite{ziegler2024measuring}. However, gains in developer speed due to AI assistance can be contingent on developer experience and task complexity, as confirmed by the controlled study by Paradis et al.~\cite{paradis2024much}, conducted at Google. 
Such performance variability across users further highlights the need for AI integration strategies that are adaptive to the needs of individual developers. The large-scale survey conducted by Sergeyuk et al.~\cite{sergeyuk2025using} showed that one significant barrier to the broader adoption of AI coding assistants is the challenge of communicating programmers' intentions to the AI assistant. Both studies indicate the need for context-sensitive user-centered AI integration strategies. To facilitate context-aware interactions, Ross et al.~\cite{ross2023programmer} introduced the Programmer's Assistant, a conversational interface grounded in the surrounding code context, enabling developers to engage in multi-turn dialogues with large language models (LLMs). 
However, the effectiveness of LLM-based tools relies on prompt formulation~\cite{sikha2023mastering, Marvin2024Prompt} (e.g., providing sufficient context in the prompts) to articulate one's intent, which poses a challenge due to the imposed cognitive load~\cite{hassan2024towards}, especially for novice programmers~\cite{shah2025students, yeh2025bridging, mailach2025interaction}. 

While previous work highlights the limitations of prompt quality that can lead to a loss of trust in AI-assisted coding, it overlooks how real-time cognitive signals can enhance AI in coding. 
We propose gaze-informed prompt engineering, which uses eye-tracking to adjust prompts based on user attention, enabling real-time, user-aligned code suggestions with minimal input. Such an approach is also a step towards eliminating the need for prompt engineering, which is a challenge to overcome, as Hassan et al.~\cite{hassan2024towards} point out, to effectively and efficiently leverage the complementary strengths of human developers and sophisticated AI systems.

\smallskip
\noindent \textbf{Code Readability and Comprehension.}
Software developers spend a significant portion of their time reading code. Brooks et al.~\cite{brooks1983towards} argued that code should be written to minimize the time it takes someone else to understand it. In line with this foundational view, in the responses of the survey Ljung et al.~\cite{ljung2022clean} conducted, software developers endorsed the principles of Clean Code, associating them with improved maintainability and efficiency. The study also revealed that developers often prioritize speed and functionality in early iterations, refining readability in later stages. 
However, as Johnson et al.~\cite{johnson2019empirical} argue, neglecting readability imposes cognitive effort on developers, which can hinder long-term code comprehension and slow down maintenance, ultimately creating productivity bottlenecks. Code comprehension is crucial for effective software maintenance.

The need to support code comprehension (e.g., improving code readability, providing information support) has significantly grown with the emergence of AI code assistants. Prather et al.~\cite{prather2024widening} observed that many novice programmers misunderstood or unquestioningly accepted AI-generated code, leading to logic errors and poor outcomes. The study by Mailach et al.~\cite{mailach2025interaction} also revealed that beginners frequently struggled to understand or adapt AI-generated code, resulting in low-quality or incomplete solutions, unless participants engaged in iterative refinement or requested clarifications.
To aid code understanding, Nam et al.~\cite{nam2024using} developed GILT, an IDE plugin that leverages OpenAI's GPT-3.5-turbo to provide explanations for highlighted sections of code. While GILT focuses on providing code explanations to aid in code understanding, our approach prioritizes enhancing code readability. Similar to our approach, using GILT does not require writing prompts; instead developers can select from high-level request options (e.g., to get a summary of highlighted code, explanations on API calls). In contrast, our proposed approach adapts to users' comprehension as they interact with the code by leveraging gaze data as a real-time signal of cognitive load and attention. 

\smallskip
\noindent \textbf{Eye Tracking in Software Engineering.} The integration of eye tracking into software engineering research has been advanced by recent methodological frameworks that highlight the value of gaze metrics for uncovering real-time comprehension patterns \cite{grabinger2025cookbook}. Eye-tracking has proven valuable for studying how developers interact with code, offering insights into debugging \cite{Jensen2010GazeStrategies, lin2015tracking}, code summarization \cite{rodeghero2014improving, karas2024tale}, and problem-solving \cite{yang2024attention}. It reveals cognitive processes behind comprehension and how expertise shapes attention patterns \cite{bednarik2004visual, Busjahn2015EyeMovements, aljehane2023studying, mohamed2025design}. Gaze has also been explored in collaborative development. In remote pair programming, shared gaze improved referential clarity and joint focus \cite{Angelo2016gazed}. Dual eye-tracking studies further show that gaze coupling indicates shared understanding and predicts collaborative success \cite{villamor2018predicting}. However, these studies primarily focus on human–human collaboration.

More recently, gaze has been applied to AI-assisted programming. Tang et al.\cite{tang2024developer} found that developers unaware of an LLM’s involvement were more likely to overlook logic flaws, illustrating automation bias. Mohamed et al.\cite{mohamed2025design} observed shifts in student attention during AI-assisted summarization. Al Haque et al.~\cite{haque2025towards} combined gaze, EEG, and IDE logs to quantify cognitive load across AI and non-AI coding tasks.
Other studies report reduced attention to AI suggestions \cite{al2022readable, mozannar2024reading}, with delayed verification increasing cognitive burden. These issues are especially pronounced for novices—Prather et al.~\cite{prather2024widening} used gaze to uncover how students with low self-efficacy were more susceptible to misleading AI output and less able to catch their own mistakes.


However, most prior work treats gaze as a retrospective signal. Obaidellah et al.’s survey \cite{obaidellah2018survey} emphasized its untapped potential for real-time interaction. Our work addresses this gap by treating gaze as an active input rather than a passive signal. We embed gaze into the prompts, allowing LLM outputs to adapt in real-time based on user attention and cognitive state—enhancing comprehension, readability, and the overall quality of AI assistance.

\begin{table*}[htbp]
    \small  
    \caption{Gaze metrics used by \gaze pipeline and to generate prompt text. The fixed prompt used in the \gazeAndText treatment also relies on the listed gaze metrics.}
    \label{tab:gaze-metrics}
    \centering
    \begin{adjustbox}{max width=\linewidth}
    \begin{tabular}{l p{9cm} >{\centering\arraybackslash}m{4.5cm}}  
        \hline
        \textbf{Gaze Metric$^{\dagger}$} & \textbf{Definition} & \textbf{Formula$^{\ddagger}$} \\
        \hline \hline
        \rowcolor{black!5}
        Mean Fixation Duration & The average duration of fixations, where each fixation duration is the difference between the timestamps of the first and last gaze points within a fixation cluster. &
        $\displaystyle \frac{1}{N} \sum_{i=1}^{N} (\text{Time}_{\text{end}, i} - \text{Time}_{\text{start}, i})$ \\

        Mean Fixation Count per Second & Total number of fixations divided by the total gaze recording time in seconds &
        $\displaystyle \frac{\text{Total Fixations}}{\text{Total Time (ms)} / 1000}$ \\

        \rowcolor{black!5}
        Mean Saccade Length & The average Euclidean distance between successive gaze points, assuming the gaze coordinates are normalized and the screen width is 1920 pixels. &
        $\displaystyle \sqrt{(x_{i} - x_{i-1})^2 + (y_{i} - y_{i-1})^2} \times 1920\ \text{px}$ \\

        Mean Pupil Dilation & The average change in pupil diameter relative to a baseline. The baseline is calculated from samples in the first 60 milliseconds of the coding session. &
        $\displaystyle \frac{1}{M} \sum_{i=1}^{M} (\text{Pupil}_i - \text{Baseline})$ \\
        \hline
        \multicolumn{3}{l}{\footnotesize \emph{($\dagger$) Mean fixation duration is in milliseconds, Mean saccade length is in pixels, and Mean pupil dilation is in millimeters.}} \\
        \multicolumn{3}{l}{\footnotesize \emph{($\ddagger$) In the formulas, $N$ is the number of total fixations and $M$ is the total number of pupil dilations.}} \\
    \end{tabular}
    \end{adjustbox}
    \end{table*}

\section{\gaze: Concept and Implementation}

To support the aims of this research, we developed \gaze, a novel software pipeline that integrates real-time eye tracking with AI-assisted refactoring to improve code readability. \gaze leverages the CodeGRITS toolkit \cite{tang2024codegrits}, a JetBrains IDE plugin for software engineering research to enable the generation of context-aware prompts for GitHub Copilot based on developers’ gaze behaviour while reading source code. The \gaze software pipeline is publicly available in the replication package \cite{GazeCopilot2025}. 

\subsection{Prompt Design}
\label{subsec:Prompt Design}
\noindent \textbf{(1) Selecting eye-tracking metrics.} 
We selected eye-tracking metrics based on prior research highlighting their relevance in distinguishing how programmers interact with code. \textit{Fixation duration}, \textit{fixation count per second}, and \textit{saccade length} capture eye movement patterns that reflect the cognitive demands of readable versus poorly structured code \cite{Jensen2010GazeStrategies}, while \textit{pupil dilation} is a well-established indicator of cognitive load during complex tasks such as code comprehension \cite{aljehane2023studying}. These metrics also help differentiate the gaze behaviour of novice and expert programmers, with novice patterns often indicating difficulty in comprehension and a greater need for assistance \cite{aljehane2023studying}.

\smallskip
\noindent \textbf{(2) Selecting eye-tracking thresholds.} 
We employed \textit{predefined thresholds} for each of the eye-tracking metrics as indicators that the developer may be struggling to comprehend the code. These thresholds are as follows: fixation duration = 241.31 ms, fixation count per second = 2.89, saccade length = 132.74 pixels, and relative pupil dilation = 0.1 mm. The threshold values were chosen based on mean values reported by Jensen et al.~\cite{Jensen2010GazeStrategies} on novice developers' gaze behavior with well-structured code \cite{Jensen2010GazeStrategies}. We used these thresholds as reference points to indicate “high” levels of each eye-tracking metric when constructing the prompt, since our pilot tests showed that Copilot does not meaningfully interpret the raw numerical values of these metrics.

To confirm that our thresholds are suitable for the demographics of our recruited participants, we compared our participants' demographics with those of the participants in the study by Jensen et al.~\cite{Jensen2010GazeStrategies} and found them to be largely similar. The participants in the referenced study~\cite{Jensen2010GazeStrategies} were novice programmers with an average experience level of 2.35 ($SD = 0.51$) on a scale from 0 to 6. Similarly, the majority of our participants rated their programming experience compared to their peers as “average” ($n = 14$). This form of self-assessment has been empirically shown to be a reliable measure of programming experience, as demonstrated by Feigenspan et al.~\cite{feigenspan2012measuring}, whose model was also used by Jensen et al.~\cite{Jensen2010GazeStrategies}.
This high similarity suggests that the thresholds chosen for our study are suitable for our recruited participants.  

\smallskip
\noindent \textbf{(3) Mapping eye-tracking metrics to cognitive states.} 
We inferred programmers' cognitive states by their gaze data while they read code containing readability challenges (i.e., messy code), referring to previous research in the literature, as follows:
\begin{itemize}[left=0pt]
    \item \textbf{Mean fixation duration:} Longer fixation durations indicate sustained attention, deeper processing, and higher cognitive effort, often due to increased task complexity or ambiguity in the visual stimulus. This may be due to code complexity or the difficulty of detecting defects, requiring greater cognitive effort to process the given task \cite{aljehane2023studying, sharafi2015eye, busjahn2011analysis}.
   \item \textbf{Mean fixation count per second:} A higher fixation count typically suggests lower efficiency, as it reflects increased visual effort to locate relevant information within an area of interest (AOI) or stimulus \cite{aljehane2023studying}. It can also indicate that the layout or content requires more scanning, suggesting cognitive strain in locating meaningful cues \cite{sharafi2015eye}.
    \item \textbf{Mean saccade length:} Shorter saccades are associated with novice behavior, while experts tend to make larger jumps between code blocks, as they can identify important sections and do not need to read the code line by line \cite{aljehane2023studying}. Prior research also shows that novice programmers read code more linearly compared to experts \cite{Busjahn2015EyeMovements}.     
    \item \textbf{Pupil dilation:} Increased pupil dilation serves as a sensitive indicator of cognitive challenge, reflecting the mental effort required when working on difficult tasks \cite{aljehane2023studying, fritz2014using}. 
\end{itemize}

\begin{figure}[h]
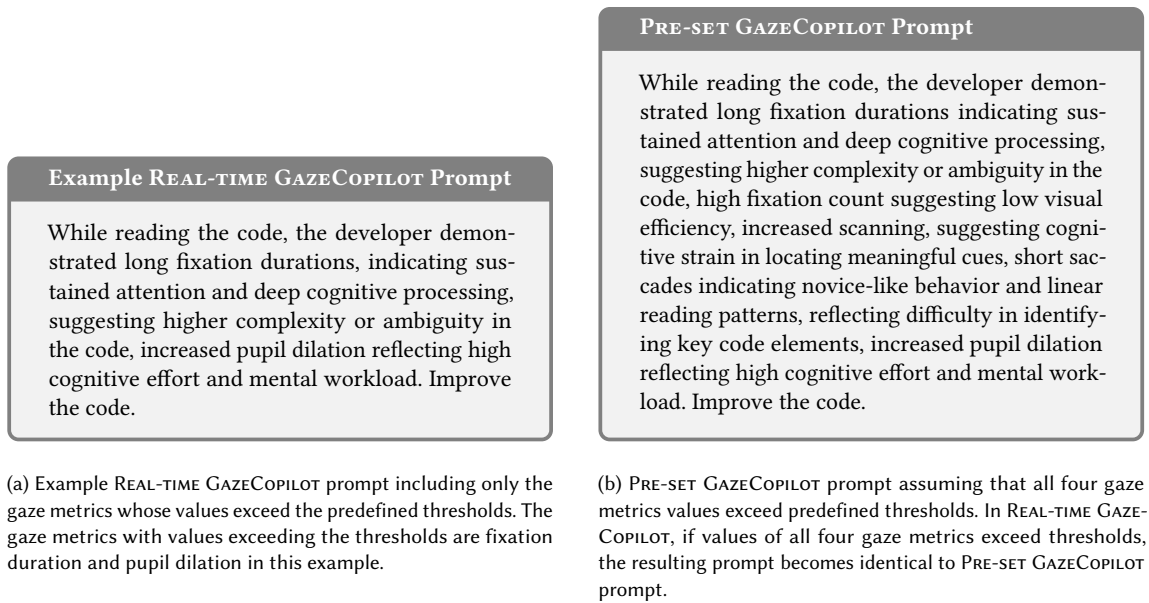

    \centering
    \begin{subfigure}[t]{0.48\textwidth}
        \begin{tcolorbox}[
            colback=black!5,
            colframe=black!50,
            title=\textbf{Example \gaze Prompt}
        ]
        While reading the code, the developer demonstrated long fixation durations, indicating sustained attention and deep cognitive processing, suggesting higher complexity or ambiguity in the code, increased pupil dilation reflecting high cognitive effort and mental workload. Improve the code.
        \end{tcolorbox}
        \caption{Example \gaze prompt including only the gaze metrics whose values exceed the predefined thresholds. The gaze metrics with values exceeding the thresholds are fixation duration and pupil dilation in this example.}
        \label{fig:example-prompt}
    \end{subfigure}
    \hfill
    \begin{subfigure}[t]{0.48\textwidth}
        \begin{tcolorbox}[
            colback=black!5,
            colframe=black!50,
            title=\textbf{\gazeAndText Prompt}
        ]
        While reading the code, the developer demonstrated long fixation durations indicating sustained attention and deep cognitive processing, suggesting higher complexity or ambiguity in the code, high fixation count suggesting low visual efficiency, increased scanning, suggesting cognitive strain in locating meaningful cues, short saccades indicating novice-like behavior and linear reading patterns, reflecting difficulty in identifying key code elements, increased pupil dilation reflecting high cognitive effort and mental workload. Improve the code.
        \end{tcolorbox}
        \caption{\gazeAndText prompt assuming that all four gaze metrics values exceed predefined thresholds. In \gaze, if values of all four gaze metrics exceed thresholds, the resulting prompt becomes identical to \gazeAndText prompt.}
        \label{fig:preset-prompt}
    \end{subfigure}
    \caption{Comparison of gaze-informed prompts used in \gaze and \gazeAndText, showing the text provided to Copilot in each treatment.}
\end{figure}

\smallskip
\noindent \textbf{(4) Generating gaze-informed natural language prompt.} 
We incorporate the eye-tracking metrics that exceed the predefined thresholds, along with the associated inferred cognitive states, into a natural language prompt.
The resulting prompt text first describes the programmers' cognitive state (while reading the messy code) based on these metrics, followed by the command: "Improve the code."
For instance, if two of the gaze metrics values (e.g., mean fixation duration and mean pupil dilation) exceed the predefined thresholds, we include these metrics along with the cognitive states associated with them as shown in Figure~\ref{fig:example-prompt}.

\smallskip
\noindent \textbf{(5) Pilot Testing of Prompt Design.}
To explore whether GitHub Copilot takes gaze metrics into account when included in the prompt, we conducted pilot tests prior to conducting the main study to assess whether Copilot's responses are influenced by the provided context about programmers' gaze behavior compared to generic instructions without the gaze data (e.g., "Improve the code").


For each gaze metric (see Table~\ref{tab:gaze-metrics}), we generated a prompt following a format similar to that shown in Figure~\ref{fig:example-prompt}. However, each prompt included only information about the programmer's cognitive state relevant to that specific gaze metric. For example, the prompt we prepared for \textbf{mean fixation duration} was: "\textit{While reading the code, the developer demonstrated \textbf{long fixation durations}, indicating sustained attention and deep cognitive processing, suggesting higher complexity or ambiguity in the code. Improve the code.}".
We also experimented with different combinations of gaze metrics to determine their effectiveness. Each generated prompt—whether based on a single metric or a combination—was then fed separately to Copilot for evaluation.



The results of our pilot testing are below. They are based on \textbf{(i)} our own qualitative evaluation of the resulting responses, and \textbf{(ii)} an examination of Copilot's rationale behind the response when prompted: "Describe how each of the gaze metrics was interpreted and mapped to specific code changes." and "Explain how the gaze metric influenced your decision to refactor the code."

\begin{itemize}[left=0pt]
    \item To address \textbf{high fixation count}, Copilot suggests more \textit{meaningful variable names} and adds \textit{concise comments}, making the code more self-explanatory and allowing users to focus on the logic rather than decoding the syntax or structure.

    \item Copilot mitigates \textbf{long fixation durations} by using \textit{descriptive} and \textit{consistent} identifiers and by simplifying logic to reduce \textit{duplicate or redundant code}, thereby lowering the cognitive load per line of code.

    \item Copilot responds to \textbf{short saccade lengths} by enforcing \textit{consistent formatting} and \textit{modular structure} by breaking down long methods into smaller,  methods that have \textit{single-responsibility}, facilitating the grasp of the relationships between code blocks.

    \item Copilot aims to reduce cognitive load indicated by \textbf{pupil dilation} by breaking down code logic, \textit{adding comments}, and avoiding ambiguous or error-prone programming patterns, such as \textit{deeply nested conditionals}, \textit{unclear variable names}, or \textit{overly complex expressions}.
\end{itemize}

\begin{figure}[h]
  \centering
  \includegraphics[width=11cm]{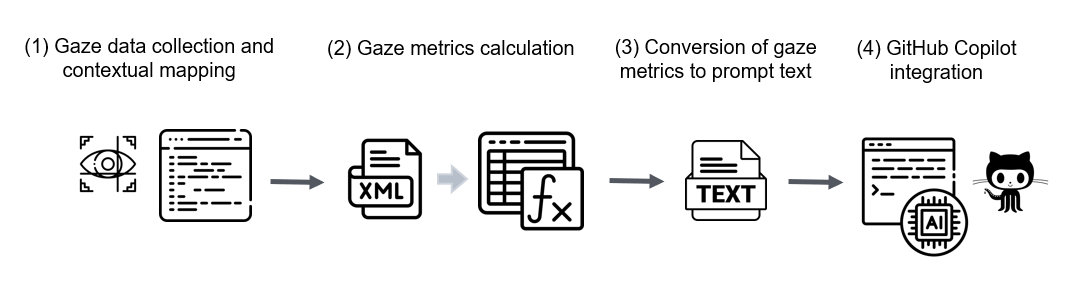}
  \caption{\gaze pipeline}
  \label{fig:GazeCopilot-pipeline}
  \vspace{-3mm}
\end{figure}

\subsection{\gaze Pipeline}
Figure~\ref{fig:GazeCopilot-pipeline} presents a high-level overview of the \gaze pipeline, which we describe below.

\smallskip
\noindent \textbf{(1) Gaze data collection and contextual mapping.}
We employed the \textit{Tobii Pro Spark} \cite{tobiispark} eye tracker, which allows for unobtrusive gaze capture during extended programming sessions, even with moderate head movements, since it is a screen-mounted (non-wearable) tracker that supports 60 Hz binocular tracking and automatic user detection. To interface with \textit{Tobii Pro Spark}, we used the \textit{Tobii Pro SDK for Python}~\cite{tobiisdk}, that provides real-time access to streamed gaze data including gaze coordinates, pupil diameter, gaze validity, and timestamps. \textit{CodeGRITS} integrates with the Tobii Pro SDK, allowing us to capture both IDE interaction events (i.e., programmers' IDE interactions and screen recording) and gaze data in real-time, and synchronize them during the programming session. 
Once gaze data is streamed via the SDK, \textit{CodeGRITS} maps each gaze point to the corresponding source code elements (i.e., file path, line, and column), synchronising it with IDE usage logs and screen recordings and creates a structured XML file with this mapping.

\smallskip
\noindent\hypertarget{gaze-metrics}{\textbf{(2) Gaze metrics calculation.}}
From the structured XML output generated by \textit{CodeGRITS}, the pipeline calculates the four core eye tracking metrics described in Table ~\ref{tab:gaze-metrics}: mean fixation duration, mean fixation count per second, mean saccade length, and mean pupil dilation. 

\smallskip
\noindent \textbf{(3) Conversion of gaze metrics to prompt text.} 
The pipeline checks whether the calculated mean values of gaze metrics exceed the \textit{predefined thresholds}. If a metric surpasses its threshold, it is included in the prompt along with the corresponding cognitive state (see Section \ref{subsec:Prompt Design}).

\smallskip
\noindent \textbf{(4) GitHub Copilot integration.} 
We chose to feed the generated prompt text into GitHub Copilot~\cite{chen2021evaluating} due to its widespread usage and seamless integration with modern IDEs. As of early 2025, over 15 million developers across more than 50,000 organizations were using GitHub Copilot~\cite{tenet2024copilotstats}. Furthermore, GitHub Copilot integrates smoothly with IntelliJ IDEA via the official JetBrains plugin, aligning with our gaze data collection setup based on CodeGRITS.

The user interaction with \systemname is similar to standard GitHub Copilot. Users trigger \systemname via a keyboard shortcut, which displays a prompt generated based on their gaze behavior, allowing them to preview the suggested code.


\section{Methodology - Evaluating \gaze}
The goal of this study is to investigate the impact of integrating real-time eye-tracking information into LLM prompts on code readability, comprehension and developers' experience in AI-assisted programming. All experiment artifacts are publicly available in the replication package \cite{GazeCopilot2025}.

\subsection{Research Questions}
We structure our study into two main research questions. 
\begin{itemize}[left=4.5mm]
    \item[\textbf{RQ1}] How does incorporating \textit{real-time} eye-tracking data into LLM prompts impact the readability and comprehension of AI-generated code?
    \item[\textbf{RQ2}] How does the inclusion of \textit{real-time} eye tracking data into LLM prompts impact the developers' experience?
\end{itemize}






\subsection{Experiment Design and Treatments}
\label{sec:experimentDesign}

The study employs a within-subjects experimental design with one independent variable: \interactionMode, which refers to the type of input prompt that is fed into GitHub Copilot. Each participant is exposed to all of the following three \interactionModes: 
\begin{itemize}[left=0pt]
\item \textbf{{\standard} (Baseline).} Standard GitHub Copilot refactors the code based on the textual prompts the participants enter. This condition serves as a baseline representation of standard AI-assisted programming without additional enhancements.

\item \textbf{{\gaze}.} This treatment converts \textbf{\textit{real-time}} eye tracking data of the participant into natural language text and adds it to the prompt and feeds it to Copilot. The wording depends on which eye-tracking metrics exceeded the thresholds (see an example in Figure~\ref{fig:example-prompt}).

\item \textbf{\gazeAndText}. This treatment serves as another control (besides \standard), allowing us to examine whether the participants preferred general improvements to the refactored code or more adaptive adjustments based on their real-time gaze behavior. The pre-set prompt assumes that all four eye-tracking metrics exceeded thresholds and is fed to Copilot (see Figure~\ref{fig:preset-prompt}).

\end{itemize}

We counter-balanced the order of \interactionModes and assigned a different code snippet to each mode to mitigate learning effects, resulting in six different sequences as shown in Table ~\ref{tab:experimentDesign}. We randomly assigned each participant to one of the six sequences. 

\begin{table}[htbp]
    \centering
    \caption{Experiment Design}
    \label{tab:experimentDesign}
    \begin{adjustbox}{max width=\linewidth}
        \begin{tabular}{l l l l}
            \hline
            & \multicolumn{3}{c}{\textbf{Code Snippets}} \\
            \cline{2-4}
            \textbf{Sequence} & \textbf{Snippet A} & \textbf{Snippet B} & \textbf{Snippet C} \\
            \hline
            \rowcolor{black!5} Sequence 1 & \standard & \gaze & \gazeAndText \\
            Sequence 2 & \standard & \gazeAndText & \gaze \\
            \rowcolor{black!5} Sequence 3 & \gaze & \standard & \gazeAndText \\
            Sequence 4 & \gaze & \gazeAndText & \standard \\
            \rowcolor{black!5} Sequence 5 & \gazeAndText & \standard & \gaze \\
            Sequence 6 & \gazeAndText & \gaze & \standard \\
            \hline
        \end{tabular}
    \end{adjustbox}
\end{table}

\begin{table*}[htbp]
    \small  
    \caption{Code readability violations we injected in the code snippets to simulate realistic code issues in our experiment. }
    \label{tab:readabilityChallenges}
        \centering \begin{adjustbox}{max width=\linewidth}
        \begin{tabular}{l l l }
            \hline
            \textbf{Readability Violation} & \textbf{Definition} & \textbf{Guidelines/Standards}\\
            \hline
            \rowcolor{gray05} Meaningless Naming & Classes, methods, and variables lack meaningful names, making the code difficult to interpret.& \cite{fowler2018refactoring, xia2017measuring, tashtoush2013impact}\\      
       
            Inconsistent Naming  & The code uses multiple naming styles inconsistently, making it harder to read and maintain. & \cite{fowler2018refactoring} \\  
      
            \rowcolor{gray05} Lack of Modular  & The system follows a monolithic design without a clear hierarchical structure or separation & \cite{fowler2018refactoring, Fang2001CodingStandard} \\
            \rowcolor{gray05} Structure        &  of concerns & \\        
          
            Lack of Structural  & The code lacks textual features that provide visual and structural cues, such as indentation,  & \cite{Jensen2010GazeStrategies} \\
            Textual Features    & spacing and consistent formatting, which guide the reader through the code's logical flow. & \\
           
            \rowcolor{gray05} Violation of the Single  & The methods are excessively long and handle multiple tasks. & \cite{fowler2018refactoring, xia2017measuring, Fang2001CodingStandard, tashtoush2013impact, crosby1990we}\\
            \rowcolor{gray05} Responsibility Principle & & \\
        
            Lack of Comments and  &  The code has minimal or no comments explaining the logic and intent behind various &  \cite{crosby1990we, Fang2001CodingStandard, fowler2018refactoring, tashtoush2013impact, xia2017measuring}\\
            Documentation &  components. & \\

            \rowcolor{gray05} Duplicate/Redundant & (Nearly) identical code appears in multiple places instead of being refactored into reusable  & \cite{fowler2018refactoring} \\
            \rowcolor{gray05} Code        & functions or classes. & \\
            High Cognitive  & The code contains numerous control flow structures, nested logic and decision points that & \cite{campbell2018cognitive}. \\
            Complexity &  increase mental effort to follow the execution flow and reason about potential outcomes. & \\ 
            \rowcolor{gray05} High Cyclomatic  & The code contains multiple decision points, branching paths, and deeply nested loops and  &\cite{mccabe1976complexity, fowler2018refactoring, tashtoush2013impact} \\
            \rowcolor{gray05} Complexity & conditionals making it harder to analyze and predict execution flows  &    \\ 
            \hline
        \end{tabular}  
        \end{adjustbox}
\end{table*}

\begin{figure*}[h]
  \centering
  \includegraphics[width=\textwidth]{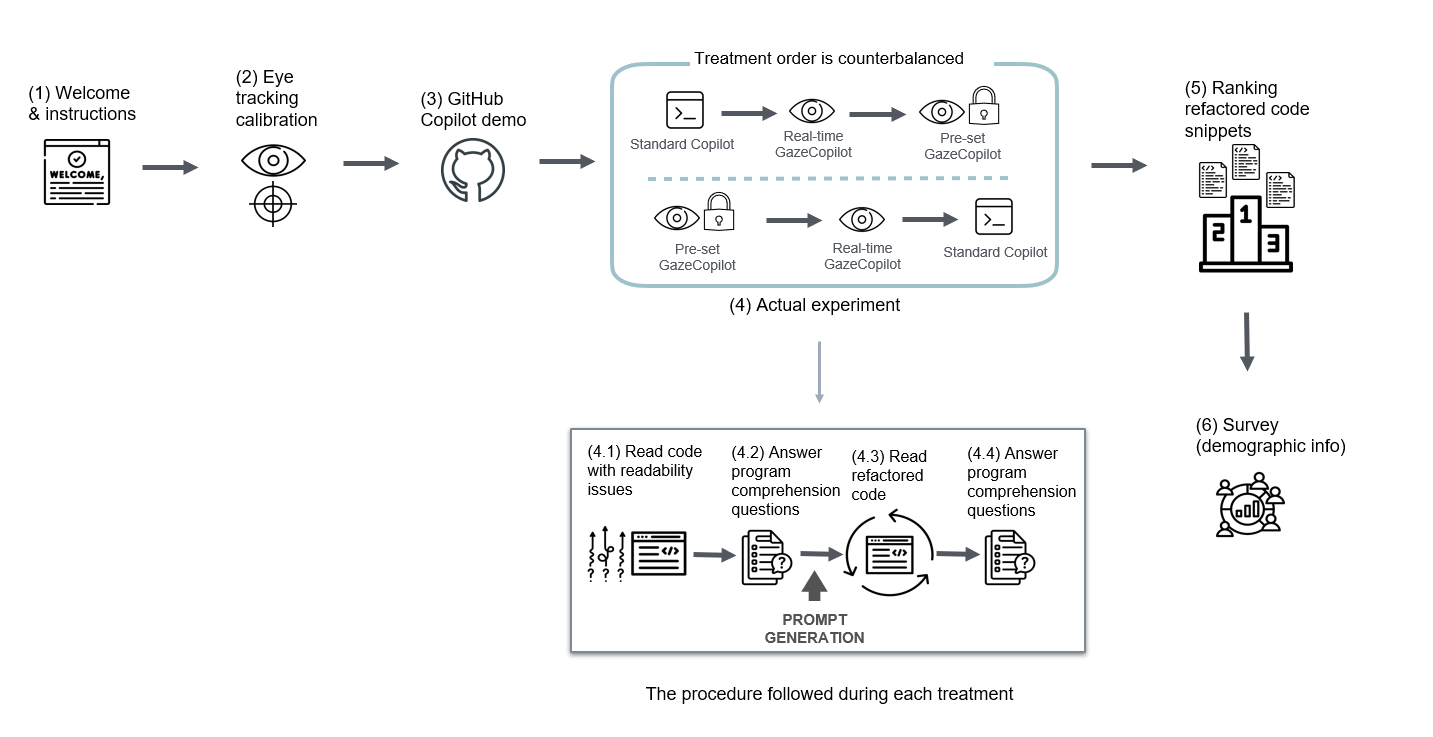}
  \caption{Experiment flow}
  \label{fig:experimentFlow}
\end{figure*}

\subsection{Experimental Procedure}
\label{subsec:experimenral-structure}
We invited participants to our lab, where the lighting was controlled. We conducted the experiment with one participant at a time. 
The experiment stages are detailed below and in Figure~\ref{fig:experimentFlow}.

\noindent \textbf{(1) Welcome and instructions.} Before the experiment began, participants received an information
sheet detailing the study’s purpose, tasks, and data collection methods. 
Participants provided informed consent before proceeding with the study. The study received ethical approval from our institution.

\noindent \textbf{(2) Eye Tracker calibration.} 
We calibrated the eye tracker for each participant using Tobii's default calibration procedure. 

\noindent \textbf{(3) GitHub Copilot demonstration.} Prior to beginning the coding tasks, participants were given a brief hands-on demonstration of GitHub Copilot to ensure they understood how to interact with the tool. The demonstration showed how to trigger suggestions using the appropriate keyboard shortcuts and preview the generated code suggestions. This step was done to ensure that participants’ familiarity with Copilot did not confound the experimental results or influence the evaluation of our system.

\noindent \textbf{(4) Actual experiment.} We assigned each participant to the three treatments (i.e., \standard, \gaze, \gazeAndText) in a counter-balanced order as mentioned in Section~\ref{sec:experimentDesign}. 

During each treatment (as shown in Figure~\ref{fig:experimentFlow}), we asked participants to read a code snippet that has readability issues (i.e., messy code). The code was displayed using the IntelliJ IDE. While participants read the messy code (i.e., in stage (4.1)) their gaze was tracked to establish a \textit{messy code reading behavior}. The participants then answered two Program Comprehension (PC) questions (stage (4.2)). 
The PC questions ask what the expected code output is, and for a high-level explanation of the code's functionality. We then asked participants to rate their confidence in their comprehension of the code. 

The code was refactored using prompts in all treatments (stage (4.3) in Figure~\ref{fig:experimentFlow}). In the \standard treatment, we instructed all participants: "You should use Github Copilot to generate an improved refactored version of the presented code. You may use however many prompts you need." All participants then created and submitted their own text prompts to Copilot. In the \gaze treatment, the prompt was automatically generated from participants' real-time gaze data. In the \gazeAndText, we used a pre-set (fixed) prompt that assumes all four eye-tracking metrics exceeded thresholds.

After refactoring, participants read the updated code, answered the same two PC questions about it, and indicated their confidence in their comprehension of the code (stage (4.4) in Figure~\ref{fig:experimentFlow}).

Following the PC questions, participants answered questions about their subjective experience, including their percieved control, sense of agency, and percieved ownership of the code. While reading the refactored code (during stage (4.3)), we also tracked the participants' gaze to examine changes in gaze behaviour compared to the \textit{messy code reading behavior}. 

\noindent \textbf{(5) Ranking refactored code snippets.} Participants then ranked the three refactored code snippets from most to least readable and justified their choices.

\noindent \textbf{(6) Demographics.} Participants then completed demographic questions, including age, gender, and highest level of education, as well as programming experience.
This information was required to help us identify the extent to which our participants represent novice developer population ~\cite{Falessi:2018} and assess the suitability of the predefined gaze metric thresholds (see Section~\ref{subsec:Prompt Design}).

\subsection{Experimental Objects}
We used three Java code snippets that are \textbf{(i)} neither too trivial nor too complicated, \textbf{(ii)} self-contained, requiring no external dependencies, and \textbf{(iii)} do not rely on special technologies or frameworks/libraries.


To introduce readability challenges, we injected code readability violations based on coding standards and refactoring guidelines in the literature (see Table~\ref{tab:readabilityChallenges}). 
To ensure similar comprehension difficulty, we used SonarQube \cite{campbell2018cognitive} to calculate cognitive complexity scores, which reflect control flow and logic complexity. The messy snippets scored 50, 36, and 40, indicating comparable cognitive complexity. According to Campbell, who developed the cognitive complexity metric \cite{campbell2018cognitive}, a threshold of 15 is recommended \cite{stackoverflow_cognitive_complexity}. All of our snippets exceed this threshold, suggesting they are sufficiently complex to require refactoring.

\subsection{Variables and Measurements}
The dependent variables are categorized into four sections:
\\
\noindent{\textbf{Code Comprehension Metrics}.} Our analyses employ comprehension metrics based on previous research~\cite{Park2024AnEyeTrackingStudy} that evaluate how source code readability impacts developers' ability to understand the code. Comprehension \textbf{accuracy} measures the correctness of participants' responses to the two PC questions, which is an ordinal value within the range $[0,2]$. Additionally, we measure comprehension \textbf{time} as the time it takes to read the code snippet and answer the PC questions. 
We also capture participants' subjective \textbf{confidence} in their code comprehension using a 5-point Likert scale, where higher scores indicate greater confidence. Participants rated their confidence both before and after refactoring—that is, for the messy and refactored versions of each code snippet.

\smallskip
\noindent{\textbf{Code Readability Metrics}.} 
At the end of the actual experiment (stage (4) in Figure~\ref{fig:experimentFlow}), we collect participants' \textbf{readability rankings} of the refactored versions of the three snippets from the most (1) to the least (3) readable.
As an objective readability metric, we use \textbf{cognitive complexity} \cite{campbell2018cognitive}, which provides a readability-oriented code complexity measure. Cognitive complexity increases as the nesting of loops (\texttt{for}, \texttt{while}) and conditionals (\texttt{if-else}, \texttt{switch}) increases. Control structures such as loops, exception handling (\texttt{try-catch}), and boolean expressions with multiple logical operators (\texttt{\&\&}, \texttt{||}) contribute to an increase in cognitive complexity. 

\smallskip
\noindent{\textbf{Eye-Tracking Metrics}.} 
We employ eye-tracking metrics (see Table~\ref{tab:gaze-metrics}) to assess gaze behaviour during the code comprehension tasks. We collect eye-tracking metrics for each participant during each \interactionMode (i.e., \standard, \gaze, and \gazeAndText). This process is performed twice: once while participants read code with readability issues (i.e. messy code), and again while they read the refactored version. 

\smallskip
\noindent{\textbf{User Perception and Agency}.} 
We evaluated participants' experience and sense of agency while using \standard, \gaze, and \gazeAndText, by conducting a separate post-task questionnaire after each \interactionMode.
We adapted the questionnaire questions from prior research on developer experience, with a particular focus on user agency and perceived system performance \cite{zhang2007evaluating, jaworski2023study}. The questionnaire consists of seven 5-point Likert scale questions where higher scores reflect more positive experiences or perceptions.

The experience questions comprise questions about \textbf{perceived understandability} (Q1), the need for \textbf{manual edits} (Q2), and code comprehension \textbf{speed} (Q7) of the refactored code, as well as the \textbf{intuitiveness} (Q6) of using each \interactionMode. 
The sense of agency questions include the extent of  \textbf{perceived control} over the refactoring process (Q3), \textbf{control satisfaction}—that is, satisfaction with the level of control (Q4)—and \textbf{code ownership}, or the extent to which the refactored code feels like one's own work (Q5).



\subsection{Statistical Analyses}

Having applied assumption checks\footnote{The results of all assumption checks conducted are in the replication package in the file \texttt{supplementary\_analyses\_results.txt} under the folder \texttt{experiment}.}, including Shapiro–{Wilk} test for normality of residuals and Mauchly’s test for sphericity, we used parametric methods where data satisfied the assumptions and non-parametric methods otherwise. 

To analyze \textbf{comprehension accuracy}, \textbf{time}, \textbf{confidence}, as well as \textbf{eye-tracking metrics}, we conducted Aligned Rank Transform (ART) ANOVA with fixed effects for \interactionMode (\standard, \gaze, and \gazeAndText), version (messy / refactored), and their interaction, and participant as a random intercept to account for within-subject variability. To investigate significant effects further, we performed post-hoc pairwise comparisons using estimated marginal means with Bonferroni correction, based on the ART-aligned linear model. To quantify the magnitude of observed differences (i.e., effect size), we calculated rank biserial ($r_{\text{rb}}$) correlations for each pairwise contrast using aligned rank-transformed data.


We used the Friedman test to analyze ordinal or non-parametric data—such as subjective \textbf{readability rankings}, the \textbf{cognitive complexity} scores of the refactored code snippets, and participants’ responses to the seven \textbf{user perception and agency} questionnaire items. When the Friedman test indicated statistically significant differences, we reported Kendall’s W as a measure of effect size. For post hoc comparisons, we used Wilcoxon signed-rank tests and calculated the corresponding rank biserial ($r_{\text{rb}}$) as an effect size estimate for each pairwise comparison.
\subsection{Pilot Runs}
We ran two pilot tests to \textbf{(i)} identify and address any technical issues, \textbf{(ii)}  evaluate the snippets' complexity, \textbf{(iii)}  assess the experiment duration, \textbf{(iv)}  verify that the study instructions and interface are clear, and \textbf{(v)} collect feedback. 
Data from these pilot runs were excluded from the final analysis. 

Following the first pilot, we revised the code snippets to ensure balanced and comparable complexity across all three tasks, and improved the questionnaire's wording for clarity. 
After the second pilot run, only minor revisions were required, and the experiment was deemed ready for the main study.

\subsection{Sample Size}
We conducted a series of \textit{a priori} power analyses using G*Power \cite{faul2009statistical} to estimate the minimum required sample size for detecting medium effects ($f = 0.25$) with $\alpha = 0.05$, power = 0.80, and three within-subject conditions, unless otherwise specified. We assumed sphericity ($\epsilon = 1$) and correlation $r = 0.6$. For nonparametric tests, Friedman and ART ANOVA, we used repeated-measures ANOVA 
as a conservative proxy for sample size estimation. Sample size estimation for repeated measures ANOVA yielded 23 participants. 
We recruited 25 participants, exceeding the required threshold to ensure adequate statistical power.


\section{Results}
\subsection{Participant Demographics}
Out of 25 participants, 16 (64\%) reported having at least two years and 10 (40\%) at least five years of programming experience. Overall, 15 participants (60\%) reported programming regularly (40\% weekly; 20\%  daily). When asked to rate their programming experience relative to their peers, 56\% rated their experience as average, 28\% as good, and 16\% as poor.

Regarding GenAI Tools, 24\% were familiar with GitHub Copilot, 68\% had used other AI code assistants (e.g., ChatGPT, Claude, Gemini, and Cursor), and some had used both Copilot and other assistants or neither.

The sample included 14 participants who identified as female (56\%) and 11 as male (44\%). As we used eye tracking, we also collected vision information: All participants reported normal or corrected-to-normal vision (e.g., using glasses or contacts).

\begin{figure}[t]
    \centering
    \includegraphics[width=0.7\textwidth]{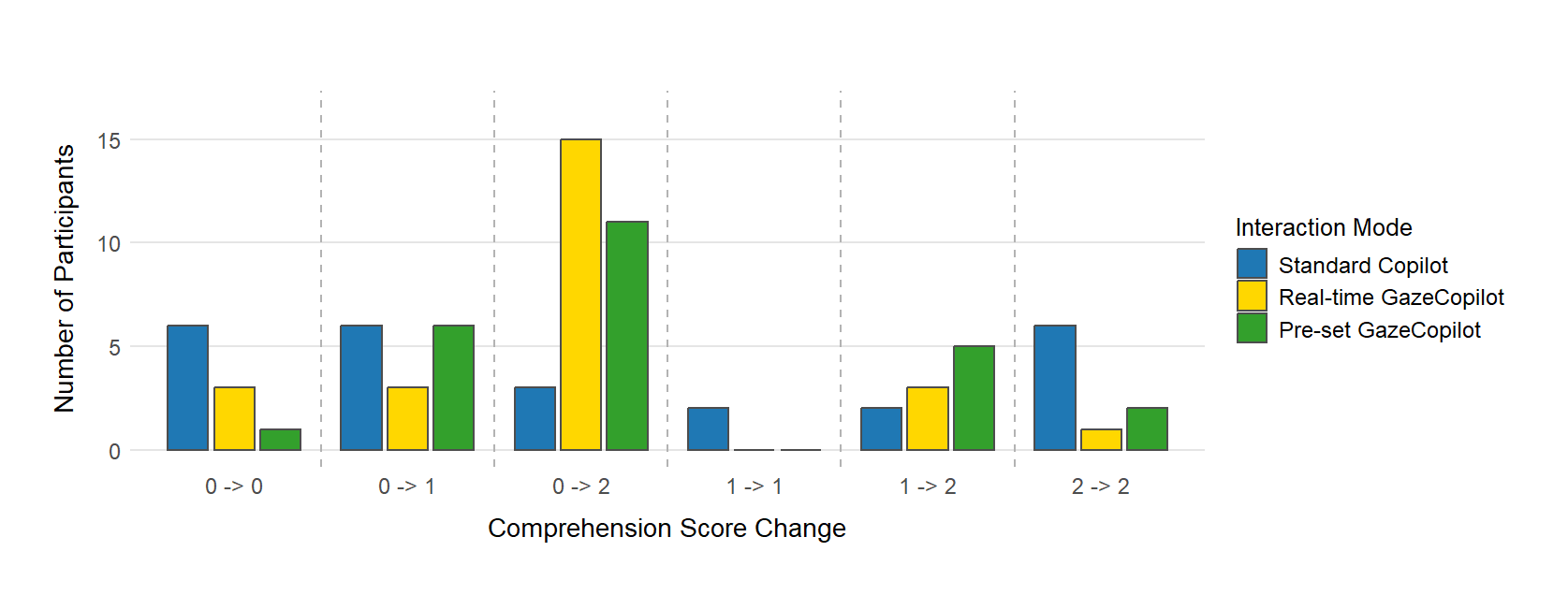}
    \caption{Comprehension score changes across \interactionModes: \standard, \gaze, and \gazeAndText(score range: 0–2, with higher scores indicating better comprehension). Each bar shows how many participants experienced a given score change $x \rightarrow y$ (e.g., 0 $\rightarrow$ 1 implies that the code comprehension score was 0 for messy code and improved to 1 for the refactored version). The \gaze \interactionMode showed the largest improvements.}
    \label{fig:comprehension-score-change}
\end{figure}

\subsection{Impact of Gaze-enhanced Prompts on Code Readability and Comprehension (RQ1)}
\label{RQ1_results}
This section presents the results of how gaze-enhanced prompts impact code readability and comprehension.

\noindent \textbf{Comprehension Accuracy.}
Figure \ref{fig:comprehension-score-change} illustrates comprehension accuracy score changes across \standard, \gaze, and \gazeAndText \interactionModes. 
It shows the number of participants whose comprehension scores changed after refactoring the messy code under each \interactionMode.
Notably, \gaze yielded the largest gains, with 15 participants improving from a comprehension score of zero to a score of two (i.e., 0 $\rightarrow$ 2), compared to 11 in the \gazeAndText and 3 in the \standard.  


To account for participants who answered correctly before the code refactoring, we calculated the improvement in comprehension score by subtracting the comprehension score for the messy version (pre-refactoring) from the score for the refactored version.
Descriptive statistics indicated the lowest mean improvement in the \standard condition ($M = 0.56$, $SD = 0.71$), with moderate gains in the \gazeAndText \interactionMode ($M = 1.32$, $SD = 0.69$), and the highest gains observed in the \gaze condition ($M = 1.44$, $SD = 0.77$).


ART ANOVA results revealed a significant main effect of \textit{version}, $F(1, 270) = 236.78$, $p < .001$, 
indicating that \textit{version} has a significant impact on comprehension accuracy, with higher scores for \textit{refactored} versions compared to the \textit{messy} ones. 
We did not find statistically significant  main effect of \interactionMode, $F(2, 270) = 0.70$, $p = .496$; however, we found a significant interaction between \interactionMode and version, $F(2, 270) = 11.46$, $p < .001$, indicating that the effect of \interactionMode on comprehension accuracy depended on whether the code version was messy or refactored. 

Post-hoc pairwise comparisons revealed that comprehension accuracy for \gaze-refactored code was significantly higher than that of \standard-refactored code ($p = .0122)$ with a moderate effect size (rank-biserial correlation $r_{rb} = 0.220$). Comprehension accuracy for \gaze-refactored code was also significantly higher than that of \gazeAndText-refactored code ($p = .0137$), with a large effect size ($r_{rb} = 0.698$).
We did not find a significant difference between \standard-refactored and \gazeAndText-refactored code ($p = 1.0000$).

In addition, post-hoc pairwise comparisons revealed a statistically significant improvement in code comprehension accuracy for \gaze-refactored ($p = .0002$) and \gazeAndText-refactored ($p = .0019$) code snippets compared to the corresponding messy versions. 




To distinguish whether prior experience with GitHub Copilot had an impact on the improvement of comprehension accuracy, we ran additional tests. Participants \textbf{familiar} with Copilot ($n = 3$) showed mean comprehension gains of \standard = $0.33$, \gaze = $0.33$, and \gazeAndText = $1.33$. Participants \textbf{unfamiliar} with Copilot ($n = 22$) showed corresponding gains of \standard = $0.59$, \gaze = $1.59$, and \gazeAndText = $1.32$. This shows that both groups benefited most from the \gaze treatment, regardless of prior Copilot experience. 


\roundedbox{
Comprehension accuracy (i.e., the correctness of responses to program comprehension questions) was significantly higher for \gaze-refactored code compared to code refactored by \standard Copilot and \gazeAndText, with a moderate and a large effect size respectively. This suggests that integrating real-time gaze data allows prompts to more effectively address developers’ comprehension challenges, which could in turn lead a more comprehensible refactored code.}

\begin{figure*}[h]
  \centering
  \includegraphics[width=\textwidth]
  {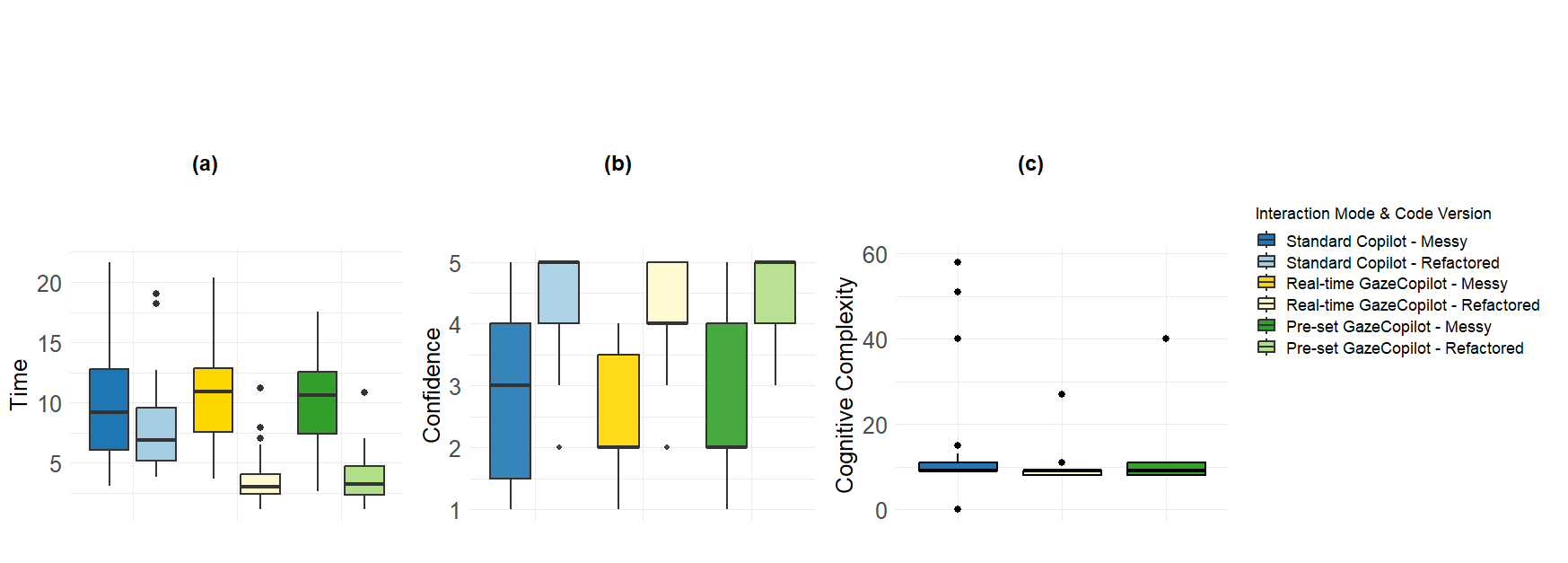}
  \caption{Box plots for (a) comprehension time in minutes and (b) comprehension confidence by \interactionMode and code snippet version (messy/refactored), and (c) refactored code snippets' cognitive complexity by \interactionMode. \gaze led to significantly faster comprehension (lower time), and the lowest mean cognitive complexity across \interactionModes.}
  \label{fig:boxplots-comprehension}
\end{figure*}

\noindent \textbf{Comprehension Time.}
ART ANOVA results did not show a significant main effect of \interactionMode, $F(2, 120) = 2.94$, $p = .057$, but showed a significant main effect of \textit{version} (messy vs. refactored code), $F(1, 120) = 114.32$, $p < .001$, as well as a significant 
interaction (between \interactionMode and \textit{version}), $F(2, 120) = 9.47$, $p < .001$. 
Follow up post-hoc analysis revealed significant differences between two pairs only: comprehension time was significantly lower in \gaze-refactored compared to \standard-refactored ($p = .0472$), and in \standard-refactored compared to \standard-messy ($p =  .0102$).
\roundedbox{For refactored code, participants completed the task faster in \gaze, with significantly shorter comprehension time than in \standard. This means that real-time gaze-informed prompts help developers process refactored code faster.}

\noindent \textbf{Comprehension Confidence.}
ART ANOVA results indicated a significant main effect of version, $F(1, 130) = 301.80$, $p < .001$, with higher comprehension confidence for refactored code compared to messy code. We did not find significant effects for \interactionMode or interaction.

\begin{figure}[t]
    \centering
    \includegraphics[width=0.4\textwidth]{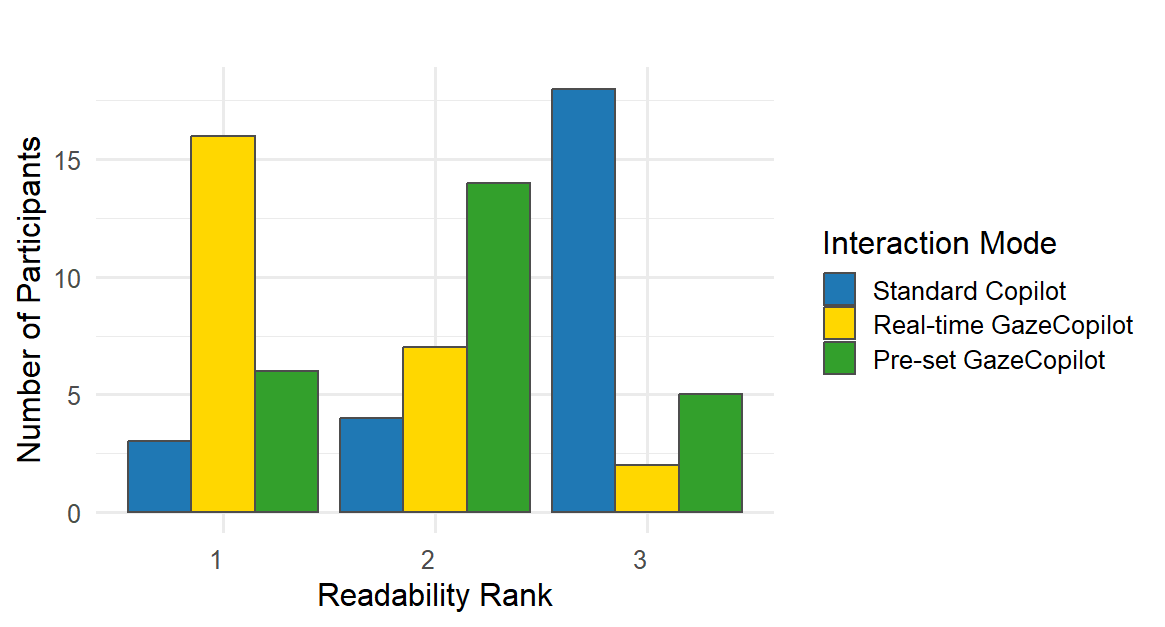}
    \caption{Distribution of readability rankings for refactored code across \interactionModes. The \gaze \interactionMode was most frequently ranked as most readable; the \standard was most often ranked as least readable.}
    \label{fig:readability-rank-distribution}
\end{figure}

\noindent \textbf{Code Readability.}
Figure~\ref{fig:readability-rank-distribution} shows that \gaze-refactored code snippets were most often rated as the most readable (64\%), whereas \standard was often ranked as least readable (72\%). \gazeAndText received intermediate rankings.

Descriptive statistics showed that the \gaze \interactionMode received the highest readability ($M = 1.44$, $Median = 1$, $SD = 0.707$), while \gazeAndText received lower rankings ($M = 1.96$, $Median = 2$, $SD = 0.676$), and \standard Copilot had the lowest readability ($M = 2.60$, $Median = 3$, $SD = 0.707$).

The Friedman test indicated a statistically significant difference across \interactionModes, $\chi^2(2) = 16.9$, $p = .000216$, with a moderate effect size (Kendall’s $W = 0.338$).

Post-hoc comparisons indicated a statistically significant difference between \standard and \gaze ($p_{\text{adj}} = .001$), with a large effect size ($r_{\text{rb}} = .7$).
We did not find a significant difference between \standard and \gazeAndText ($p_{\text{adj}} = .071$) or between \gaze and \gazeAndText ($p_{\text{adj}} = .106$).


\roundedbox{Participants ranked refactored code in \gaze as more readable than in \standard with statistical significance and a large effect size. We did not find a significant difference between \standard and \gazeAndText or between \gaze and \gazeAndText. This means that code refactored through real-time gaze-informed prompts is perceived by developers as more readable.}

\noindent \textbf{Code Cognitive Complexity.}
Descriptive statistics show that the mean cognitive complexity was lowest (i.e., best) in code refactored in \gaze ($M = 9.64$, $SD = 3.76$) , followed by \gazeAndText ($M = 10.8$, $SD = 6.22$), and highest in \standard ($M = 14$, $SD = 14$).

\roundedbox{Refactored code in the \gaze \interactionMode had lower (i.e. better) mean cognitive complexity than in \standard and \gazeAndText.}

\noindent \textbf{Eye-Tracking Metrics.} 
ART ANOVA showed significant main effects of code version (messy vs. refactored) on fixation duration ($F(1, 103.67) = 16.35$, $p < .001$), fixation count ($F(1, 104.21) = 26.37$, $p < .001$), and pupil dilation ($F(1, 107.44) = 22.62$, $p < .001$). The main effect of version on saccade length was marginally non-significant ($F(1, 103.43) = 3.28$, $p = 0.073$).

These results support our methodology of inferring code comprehension difficulties based on gaze data, as variations in these eye-tracking metrics reflect differences in cognitive processing demands across code versions (messy vs refactored).

\roundedbox{Code version (messy vs. refactored) had a significant effect on fixation duration, fixation count, and pupil dilation, with the messy version indicating higher cognitive demands. This supports our methodology, showing that gaze data reflects differences in cognitive processing and comprehension difficulties when comparing messy and refactored code.}

\noindent \textbf{Impact of Real-time Gaze Signals on Refactoring Outcomes.}
Our results highlight the importance of tailoring AI-assisted refactoring to developers’ real-time needs rather than relying on static prompts. The \gaze approach enables selective refactoring by adapting its refactoring strategy to the developer’s comprehension difficulties inferred from gaze behaviour. This minimizes redundant edits and reduces the effort required to revise code that the developer already understands.

In contrast, we applied the same hard-coded prompt for all participants in \gazeAndText, assuming that all eye-tracking metrics exceeded thresholds and thus indicating maximum need for assistance. In \gazeAndText, we found that Copilot's refactoring strategy was identical for 23 out of 25 participants, overgeneralizing refactoring and addressing more issues than the developer actually struggled with. Such excessive refactoring is not necessarily beneficial, as modifying code the developer already understands can introduce unnecessary changes and add cognitive and maintenance overhead without proportional gains~\cite{kim2014empirical, Buschmann2011, pinto2021cognitive, sellitto2022toward}.


This is further supported by the analysis of comprehension and readability metrics, which shows that \gaze-refactored code significantly outperformed \gazeAndText-refactored code in comprehension accuracy, and was more frequently rated as more readable, as detailed in Section~\ref{RQ1_results}.

\roundedbox{\gaze-refactored code significantly outperformed \gazeAndText-refactored code in \textbf{comprehension accuracy} and was more frequently rated as more \textbf{readable}. This highlights that personalising refactoring to developers’ real-time gaze behaviour (as done in \gaze) is more effective than applying generic or overgeneralised improvements that do not rely on real-time gaze data (as done in \gazeAndText), which in turn reduces unnecessary changes and the cognitive overhead required to review unneeded modifications.}




\subsection{Programmers' Perception and Agency (RQ2)}
Friedman test results did not indicate statistically significant differences between conditions for any of the user-perception questions.

\begin{figure*}[h]
\centering
\includegraphics[width=\textwidth]{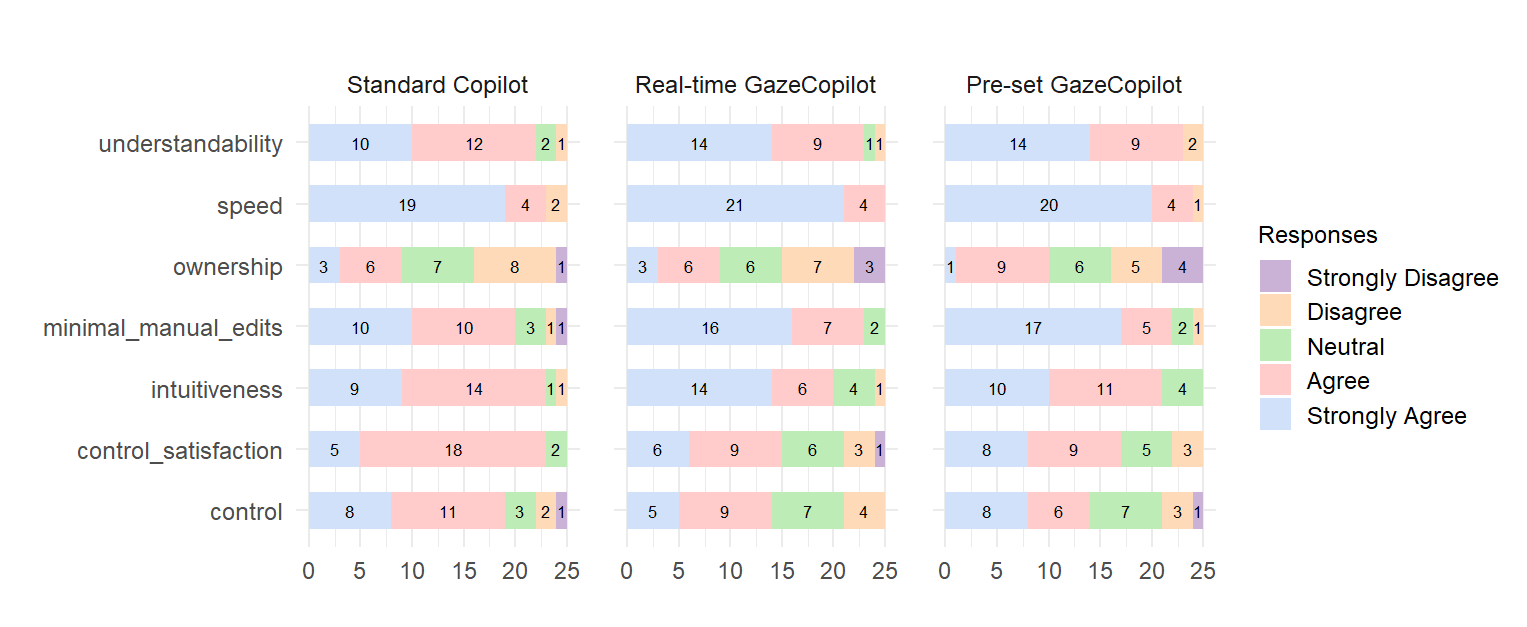}
\caption{Participants' responses to the survey questions on developers' experience and sense of agency. 
We did not find statistical differences between conditions for any of the questions.}
\label{fig:barcharts-user-perception}
\end{figure*}

Figure~\ref{fig:barcharts-user-perception} shows the distribution of participants' answers. This suggests that the \gaze prompt might be a promising addition to AI-coding assistants without negatively impacting key aspects of the developer's experience. 
One possible disadvantage of automatically generated prompts in the gaze-enhanced \interactionModes could have been the loss of the sense of agency of the programmers. We found no statistically significant differences in perceived \textbf{control}, \textbf{control satisfaction}, \textbf{intuitiveness}, \textbf{ownership}, or \textbf{speed} across the three \interactionModes. This means there is no evidence that any of the \interactionModes was perceived as more or less intuitive, controllable, or effective than the others.
\roundedbox{
We did not find evidence of differences in perceived \textbf{control}, \textbf{control satisfaction}, \textbf{intuitiveness}, \textbf{ownership}, or \textbf{speed} between the interaction modes. This suggest that \gaze-prompts might be included to AI-coding assistants without negatively impacting key aspects of developers' experience, yet such integration needs further investigation.
}

\subsection{Analysis of User Generated Prompts in \standard}
To evaluate the quality of the prompts created by participants in the \standard treatment, we sought expert feedback from two experienced software developers with 7 and 10 years of professional programming experience, respectively. Both experts code daily in professional software development environments and have experience with AI coding assistants. Each expert was asked to rate the extent to which they found each participant-generated prompt meaningful for achieving the code refactoring task, using a 5-point Likert scale (1 = Strongly Disagree, 5 = Strongly Agree). The aggregated ratings ($M = 3.06$, $Median = 3$) suggest that, overall, experts perceived the prompts as neutral in quality. 

These results indicate that the prompts generated by our participants are not meaningless and reflect the typical quality of prompts generated by developers, who often struggle to craft effective prompts, as noted in prior work \cite{sergeyuk2025using, nguyen2024beginning, liang2024large, feldman2024non}. This further highlights the need for our real-time gaze-informed prompting approach, which significantly outperformed user-generated prompts in comprehension accuracy, comprehension time, and perceived code readability, as shown in Section \ref{RQ1_results}.
The user prompts and the expert ratings are publicly available in our replication package \cite{GazeCopilot2025}).


\section{Threats to Validity}  
We recognize potential threats to the validity of our findings and describe the steps we took to mitigate them throughout the study design and execution.

\noindent \textbf{Internal validity.}  
To minimize learning effects and ordering biases, we implemented a counter-balanced within-subjects design. The order of the three \interactionModes (\standard, \gaze, and \gazeAndText) and the corresponding code snippets was counter-balanced across participants, resulting in six unique sequences (Table~\ref{tab:experimentDesign}).
This design choice helped mitigate potential confounding factors such as code-specific difficulty, task order effects, or participant fatigue. We also controlled for environmental factors by conducting all sessions in the same lab, using identical hardware, setup, and ambient lighting to ensure consistent eye-tracking accuracy.

\noindent \textbf{External validity.}  
To enhance generalizability and ensure a diverse and representative sample, the majority of participants (64\%) had at least two years of programming experience, with (40\%) reporting five or more years. Additionally, 40\% coded on a weekly basis and 20\% on a daily basis. This reflects a range of programming experience levels, enhancing the potential applicability of the study's results across the developer population.

To mitigate potential confounding factors, we employed three self-contained code snippets for our study tasks. While real-world code is typically more diverse, this controlled design minimized the influence of participants’ familiarity with specific technologies, exposure to external dependencies, and fatigue, ensuring that observed effects could be attributed solely to the experimental conditions. Moreover, it allowed us to adhere to the study duration limits set by our institution's Ethics Board, further safeguarding participants’ well-being.


\noindent \textbf{Construct validity.}  
We used established proxies for comprehension (e.g., PC questions), readability (e.g., cognitive complexity and rankings), and cognitive effort (e.g., gaze data). While we acknowledge that these proxies cannot fully capture the complexity of developers’ cognitive processes, our use of validated measures and controlled conditions helps ensure that the interpretations remain as robust and meaningful as possible.

\section{Discussion and Future Work}

\noindent \textbf{Integrating \gaze into Development Environments.}
\systemname provides a lightweight, IDE-integrated way to ease the cognitive load of manual prompt engineering, offering implicit support during code reading or refactoring. The system can be deployed in real-world applications as eye-tracking technology is advancing and becoming more accessible. 
Recent advances in visual computing allow accurate eye tracking using mobile front-facing cameras \cite{lai2025trends} and integrated webcams \cite{wisiecka2022comparison}.

This growing availability of eye-tracking technology makes it feasible to directly integrate gaze-informed refactoring into real-world development environments.
\gaze can monitor developers’ eye movements as they inspect complex or unclear code segments, automatically identifying areas that may cause cognitive strain or confusion. The system then generates targeted AI-assisted refactoring suggestions to improve readability without requiring explicit textual prompts. This capability allows developers to more quickly read, comprehend, and review code.
Further development of GazeCopilot to include features such as explaining code sections based on developers’ gaze behaviour can aid developers in debugging, refactoring, and onboarding processess. Future work should explore long-term use in real-world settings and scalability to large codebases and teams.

\smallskip

\noindent \textbf{Benefits of Real-Time Gaze Data for Software Development. } 
As LLMs have become integral to software practitioners' daily development activities, the workload for Quality Assurance (QA) activities (e.g., code review, bug triaging, testing) has increased significantly. 
According to a survey conducted with 500 software practitioners, 67\% spend more time debugging AI-generated code and 68\% spend more time fixing AI-related security vulnerabilities~\cite{Vizard2025AISurveyBadCode}. Reducing the workload on QA activities requires detecting and fixing as many issues in the code base (which mostly contains human-written code) as possible at the earliest stages of the software development cycle. Improved code readability can facilitate better code comprehension, helping developers spot issues (e.g., bugs, security vulnerabilities) and fix them early. Monitoring developers' real-time gaze data can provide insights into which aspects of the code base need refactoring to improve code readability while avoiding excessive refactoring. Moreover, improving code readability benefits software maintainability. Due to developer turnover, it is essential that not only code owners but also other developers in organisations, including incoming software engineers, can understand the code base. Improving the code base readability also supports onboarding new software engineers. 

\smallskip
\noindent \textbf{Privacy and Agency.} 
The integration of gaze-informed systems into software development workflows introduces critical considerations around privacy, transparency, and user agency. Prior work has shown that eye-tracking data, while valuable for modeling cognitive states and supporting adaptive interfaces, can inadvertently reveal sensitive information about users’ intentions, emotional states, and even task content~\cite{Alsakar2025PrivacyEyeTracking, Khamis2018PrivacyImplicationsEyeTracking}. These risks are amplified in professional programming contexts, where gaze traces may expose proprietary code or reveal aspects of developers’ problem-solving strategies~\cite{Bednarik2004DebuggingGaze, Busjahn2015EyeMovements}. As such, ensuring privacy-preserving and user-controllable deployment of gaze-informed AI assistants is essential.

To uphold developers’ autonomy and trust, gaze-informed coding tools should prioritize local data processing and explicit user consent mechanisms. Local computation mitigates risks associated with transmitting sensitive gaze data to external servers, aligning with privacy-by-design principles in physiological computing~\cite{Alsakar2025PrivacyEyeTracking, Khamis2018PrivacyImplicationsEyeTracking}. Moreover, user agency can be preserved by incorporating features such as session-level toggles—allowing developers to activate or deactivate gaze tracking at will—and transparent logging interfaces that clearly communicate when, how, and why gaze data are collected and processed~\cite{Stumpf2020UserAgencyAI, DAngelo2016GazedConfused}. Such design affordances align with HCI guidelines emphasizing explainability and informed control in adaptive systems.

Balancing automation with respect for user autonomy thus requires that gaze-informed developer tools remain user-centered and consent-aware. Transparency in data handling and fine-grained control over sensing modalities are not only ethical imperatives but also practical enablers of user acceptance and long-term adoption. By embedding these safeguards, systems like \gaze can leverage the cognitive benefits of gaze-based interaction while maintaining developers’ privacy, control, and sense of ownership over their work.

\smallskip
\noindent \textbf{Enabling Multi-modal AI Coding Assistants.} 
This work repositions gaze from a passive observational tool into an active, interpretable signal for real-time AI adaptation. By leveraging gaze as a cognitive input channel, our study highlights the potential of multi-modal coding assistants that sense and respond to a developer’s cognitive, behavioral, and affective states. Such systems could integrate complementary physiological signals, such as electroencephalography (EEG), galvanic skin response (GSR), or heart rate variability (HRV) to capture mental workload, stress, and engagement more comprehensively~\cite{Zander2011BCIAdaptiveSystems, Cowley2016PsychophysiologicalGaming} 

Future adaptive AI-assisted development environments can achieve greater robustness and personalisation by incorporating multiple modalities. For instance, EEG and eye tracking have been used jointly to detect cognitive load and attention shifts in real time~\cite {Kosch2018EEGEyeLoad, Eivazi2017EEGEyeProgramming}, enabling interfaces that dynamically adapt to users’ mental states. Similarly, multi-modal sensing has been shown to improve the detection of frustration and focus loss during complex problem-solving tasks~\cite{Cowley2016PsychophysiologicalGaming, Jraidi2013FrustrationDetection}. Translating these findings into software engineering and also referring to EEG and eye-tracking studies in software engineering~\cite{Peitek2022CognitiveLoadEEGEye, Fritz2014MeasuringProgrammerComprehension, Camilleri2022CognitiveLoadTesting}, could help design and develop systems that modulate the level of AI assistance, offering more explanations, detailed refactorings, or reduced intrusiveness—depending on the developer’s current cognitive condition.

\smallskip
\noindent \textbf{Towards Intent-Driven Software Development.} 
Our work contributes to the emerging paradigm of intent-driven software development by advancing cognitively aware, adaptive tools that respond to developers’ mental states and goals. SE 3.0 \cite{hassan2024towards} envisions AI collaborators capable of understanding developer intent and adhering to software engineering principles, while vibe coding \cite{vibecoding} emphasizes fluid, natural language interaction and real-time contextual awareness rather than rigid prompt structures. Building on these ideas, \textit{GazeCopilot} extends the notion of intent inference by using gaze as an implicit signal of cognitive state, enabling more responsive and context-aware human–AI collaboration.

In this broader landscape of AI-assisted programming, the integration of multi-modal signals, such as gaze, EEG, and interaction logs offers new opportunities for intent recognition and adaptive collaboration strategies~\cite{Hassan2024SE3Vision, Cibulski2023CognitivelyAwareIDE}. Such signals can help an assistant discern when a developer is uncertain, fatigued, or deeply engaged, and dynamically tailor the timing, modality, or granularity of its support. This kind of adaptive responsiveness aligns with established principles of human–AI teaming, which emphasize fluid coordination, mutual awareness, and calibrated trust~\cite{deVisser2020TrustHumanAITeaming}.

\section{Conclusion}
We introduce and evaluate \gaze, a novel system that integrates real-time eye-tracking data into AI-assisted prompt design to enhance code readability and comprehension. By leveraging gaze metrics such as fixation duration and pupil dilation, \gaze generates prompts that are aligned with the user's cognitive state. Our results show that real-time gaze-informed prompts significantly improve comprehension accuracy, comprehension time, and perceived code readability compared to standard text-based prompts. We demonstrate the potential of embedding implicit physiological signals to support more effective human-AI coding collaboration, especially for developers who struggle to express their intent.


\bibliographystyle{ACM-Reference-Format}
\bibliography{acmart}

\appendix

\end{document}